\documentclass{article}

\usepackage[preprint,nonatbib]{neurips_2021}

\usepackage[utf8]{inputenc} % allow utf-8 input
\usepackage[T1]{fontenc}    % use 8-bit T1 fonts
\usepackage{hyperref}       % hyperlinks
\usepackage{url}            % simple URL typesetting
\usepackage{booktabs}       % professional-quality tables
\usepackage{amsfonts}       % blackboard math symbols
\usepackage{nicefrac}       % compact symbols for 1/2, etc.
\usepackage{microtype}      % microtypography
\usepackage{xcolor}         % colors
\usepackage[pdftex]{graphicx}
\usepackage{epstopdf}
\usepackage{multirow}
\usepackage{amssymb}

\usepackage{bm}
\usepackage{amsmath}
\usepackage[export]{adjustbox} 
\usepackage[affil-it]{authblk}

\usepackage{textcomp}

\newcommand{\etal}{\textit{et al.}}

\title{A Deep Variational Bayesian Framework for \\ Blind Image Deblurring}

\author{%
  Hui Wang,\quad Zongsheng Yue,\quad Qian Zhao\thanks{Corresponding author.},\quad Deyu Meng \\
  School of Mathematics and Statistics, Xi'an Jiaotong University\\
  \texttt{huiwang2019@stu.xjtu.edu.cn, zsyzam@gmail.com,\\ \{timmy.zhaoqian, dymeng\}@xjtu.edu.cn} \\
}

\begin{document}

\maketitle

\begin{abstract}
Blind image deblurring is an important yet very challenging problem in low-level vision. Traditional optimization based methods generally formulate this task as a maximum-a-posteriori estimation or variational inference problem, whose performance highly relies on the handcraft priors for both the latent image and the blur kernel. In contrast, recent deep learning methods generally learn, from a large collection of training images, deep neural networks (DNNs) directly mapping the blurry image to the clean one or to the blur kernel, paying less attention to the physical degradation process of the blurry image. In this paper, we present a deep variational Bayesian framework for blind image deblurring. Under this framework, the posterior of the latent clean image and blur kernel can be jointly estimated in an amortized inference fashion with DNNs, and the involved inference DNNs can be trained by fully considering the physical blur model, together with the supervision of data driven priors for the clean image and blur kernel, which is naturally led to by the evidence lower bound objective. Comprehensive experiments are conducted to substantiate the effectiveness of the proposed framework. The results show that it can not only achieve a promising performance with relatively simple networks, but also enhance the performance of existing DNNs for deblurring.
\end{abstract}

\section{Introduction}\label{sec:introduction}

Image deblurring is an important problem in the field of image restoration and has attracted wide attention with a long history \cite{josa1972_richardson_bayesian_deblurring, siggraph2006_fergus_removing_shake, tip1998_chan1998_tv_deblur}. The goal of deblurring is to recover the underlying clean image from an observed blurry one. Such a blurry image can be resulted by different reasons, such as defocusing and camera shaking, which can in general be modeled as (assuming the blur process is uniform and spatially invariant)
\begin{equation}\label{blurmodel}
\bm{y} = \bm{h} \ast \bm{z} + \bm{n},
\end{equation}
where $\bm{y}$ is the blurry image, $\bm{z}$ is the corresponding clean one, $\bm{h}$ is the blur kernel (point spread function), $\bm{n}$ is the noise, often modeled as additive white Gaussian noise, and $\ast$ denotes the 2D convolution operator. When the blur kernel $\bm{h}$ is unknown, we need to simultaneously estimate $\bm{h}$ and $\bm{z}$ solely from the given observation $\bm{y}$, which is commonly referred as blind image deblurring or blind deconvolution. It is a highly ill-posed inverse problem, since there can be different pairs of $\bm{z}$ and $\bm{h}$ explaining one same blurred image $\bm{y}$.

Many different methods have been proposed to address this problem, which can be mainly categorized into two groups, i.e., optimization-based methods and deep learning ones. Traditional optimization-based methods, from the probabilistic perspective, generally formulate the image deblurring task as a maximum-a-posteriori (MAP) estimation problem \cite{cvpr2011_krishnan_normalized_sparsity_prior, tog2008_shan_high_quality, cvpr2013_xu_l0norm, cvpr2014_pan_l0text, cvpr2016_pan_dark_channel, eccv2014_michaeli_2014_patch_prior, nips2009_krishnan_hyper_laplacian} or Bayesian posterior inference (specifically variational inference (VI)) one \cite{siggraph2006_fergus_removing_shake, tip2015_zhou_variational_dirichlet_deblur, book2017_bishop_review}, by properly designing priors to both the to-be-estimated image \cite{cvpr2013_xu_l0norm, cvpr2014_pan_l0text, cvpr2016_pan_dark_channel, nips2009_krishnan_hyper_laplacian, siggraph2006_fergus_removing_shake, eccv2014_michaeli_2014_patch_prior} and blur kernel \cite{icassp1997_molina_dirichlet_deblur, tip2015_zhou_variational_dirichlet_deblur}. However, the manually designed priors are still subjective and may not precisely characterize the intrinsic distributions of natural images and blur kernels in real world, limiting the performance of a deblurring algorithm. Besides, solving the complex optimization usually causes heavy computational cost, which is always unaffordable in real applications. Recently, motivated by their great success in computer vision, especially for other image restoration problems, \cite{pami2015_dong_cnn_sr, tip2017_zhang_dncnn}, deep learning methods have also attracted increasing attention in image deblurring research. 
%\cite{cvpr15_sun_learning_kernel, bmvc15_hradis_cnn_deblur, cvpr17_nah_deepdeblur, pami15_schuler_learning_deblur}.
Early attempts along this line used deep neural networks (DNNs) as a component of traditional methods, e.g., estimating the blur kernel \cite{cvpr15_sun_learning_kernel, cvpr17_gong_motion_flow, pami15_schuler_learning_deblur, eccv16_chakrabarti_deblur}. Later, learning an explicit mapping from a blurred image to its deblurred one in an end-to-end fashion 
%from a collection of blurry-clean image pairs 
has become a dominant methodology \cite{bmvc15_hradis_cnn_deblur, cvpr17_nah_deepdeblur, cvpr18_tao_srn}. This approach is especially superior in extracting underlying knowledge from large datasets and fast deblurring for test images.

Existing deep deblurring methods mainly focus on designing fancy network architectures, while pay less attention to make full use of the information delivered by physical blur model \eqref{blurmodel}, which limits further performance improvement for them. In contrast, \eqref{blurmodel} plays a crucial role for achieving a promising performance in optimization based methods, especially Bayesian ones \cite{siggraph2006_fergus_removing_shake, cvpr2009_levin_understanding_deblurring, pami2020_pan_physics_gan}. Based on these observations, and also inspired by recent advances in deep Bayesian modeling for image restoration \cite{nips2019_yue_vdn}, we propose a deep variational Bayesian framework for training deblurring neural networks in the blind setting, aiming at simultaneously estimating the clean image and blur kernel. Specifically, under the generative model of blurry image, we construct the variational distributions parameterized by DNNs to approximate the true posteriors of both the latent clean image and blur kernel, and learn the network parameters by optimizing the evidence lower bound (ELBO). After training, the inference networks can then be directly used to approximate the corresponding posteriors of clean image and blur kernel in an amortized way, efficiently from an observed blurry image.

The main advantages of the proposed method can be summarized as follows. First, by virtue of the Bayesian generative model, the physical blur model \eqref{blurmodel} can be naturally encoded as the likelihood term in the ELBO to guide learning the inference networks. Therefore, the inference networks for the clean image and blur kernel can both be jointly trained under the supervision of data driven priors, and interact with each other due to the likelihood. Second, the proposed framework provides, in principle, a natural unification of previous methodologies that consider kernel information for training deblurring DNNs, in a concise Bayesian way, and thus is expected to achieve a better performance. Third, our framework can also be straightforwardly incorporated with any existing end-to-end DNNs to enhance their performance. All of these advantages have been substantiated by our experiments.

The remainder of this paper is organized as follows: Section \ref{sec:related_work} introduces the related studies. Section \ref{sec:method} presents the proposed deep Bayesian framework, including Bayesian generative model, deep VI procedure and network architectures. Section \ref{sec:experiments} shows the experimental results and analyses. Final conclusion is made in Section \ref{sec:conclusion}.

\section{Related Work}\label{sec:related_work}

%In this section, we review two main categories of blind deblurring methods, i.e., optimization based methods and deep learning based methods.

\textbf{Optimization based methods} 
The main focus along this research line is to design priors that can precisely characterize the intrinsic structures of clean images and blur kernels, in order to make the ill-posed blind deblurring problem solvable.
%Since blind image deblurring is highly ill-posed, it is necessary to design appropriate priors to constrain the solution space of the latent clean image and blur kernel. Optimization-based deblurring methods mainly focus on designing proper priors for latent clean images and blur kernels. 
A series of priors have been developed for clean images, including total variation \cite{tip1998_chan1998_tv_deblur}, sparsity of gradients \cite{siggraph2006_fergus_removing_shake, tog2008_shan_high_quality}, hyper-Laplacian \cite{nips2009_krishnan_hyper_laplacian}, $L_0$-norm prior \cite{cvpr2013_xu_l0norm, cvpr2014_pan_l0text}, internal patch prior \cite{eccv2014_michaeli_2014_patch_prior} and dark channel prior \cite{cvpr2016_pan_dark_channel}. For blur kernels, the aforementioned priors can still be applied, while there are also some specifically designed priors, e.g., Dirichlet prior \cite{icassp1997_molina_dirichlet_deblur, tip2015_zhou_variational_dirichlet_deblur} and piecewise-linear prior {\cite{tip2014_oh_piecewise_prior}}.
After specifying the priors, such a blind image deblurring problem can be formulated as a MAP estimation\cite{tog2008_shan_high_quality, cvpr2013_xu_l0norm, cvpr2014_pan_l0text, cvpr2016_pan_dark_channel, eccv2014_michaeli_2014_patch_prior, nips2009_krishnan_hyper_laplacian} or Bayesian posterior inference (more specifically VI) \cite{siggraph2006_fergus_removing_shake, tip2015_zhou_variational_dirichlet_deblur} problem, and solved by off-the-shelf optimization tools.

\textbf{Deep learning based methods}
Having witnessed the great success in other fields, deep learning has also been investigated in dealing with the image deblurring task. Early methods treat DNNs as a part of traditional optimization based ones, by virtue of their prediction ability. For example, Schuler \etal \cite{pami15_schuler_learning_deblur} designed a convolutional neural network (CNN) to simulate iterative optimization for image deblurring. Sun \etal \cite{cvpr15_sun_learning_kernel} used a CNN to classify the blur kernel in patch-level. Gong \etal \cite{cvpr17_gong_motion_flow} went further to predict the motion flow in a pixel-level. Chakrabarti \cite{eccv16_chakrabarti_deblur} used CNN to predict the kernel in the frequency domain. It is then gradually popular to directly parameterize the mapping from a blurry image and its clean one with CNNs in a kernel free paradigm, and then learn the parameters by a large collection of blurry-clean image pairs. Hradiš \etal \cite{bmvc15_hradis_cnn_deblur} directly predicted latent clean images from observed blurred images by a fully CNN. Nah \etal \cite{cvpr17_nah_deepdeblur} introduced a multi-scale CNN to directly recover the latent clean image.
%and constructed a realistic ground-truth blur dataset for training. 
Tao \etal \cite{cvpr18_tao_srn} improved the architecture by using a scale-recurrent network and a ConvLSTM \cite{nips2015_shi_convlstm} layer. The generative adversarial network (GAN) \cite{nips2014_goodfellow_gan} is also applied to image deblurring, e.g., Kupyn \etal \cite{cvpr2018_kupyn_deblurgan} utilized the Wasserstein GAN \cite{icml2017_arjovsky_wgan} to learn a generator that restores the latent image. Recently, there are some studies that borrow fundamental principles from traditional methods. Pan \etal \cite{pami2020_pan_physics_gan} introduce the consistency loss motivated from the likelihood in traditional methods. Cai \etal \cite{tip2020_cai_dcp_nn} embed the dark and bright channel prior to a deep multi-scale neural network. Ren \etal \cite{cvpr2020_ren_dip_deblur} tried to directly solve the traditional blind deblurring optimization, while constrain the solution space by deep image priors \cite{cvpr2018_ulyanov_dip}.

\section{Deep Variational Bayesian Framework for Blind Image Deblurring}\label{sec:method}
%In this section, we present our amortized variational inference framework for blind image deblurring. The goal of our method is to efficiently infer the posterior of latent clean image and the blur kernel from the observed blurred image. We will show our proposed method can take advantages of both optimization-based methods and deep learning methods.

Let $\mathcal{D}=\{ \bm{x}^{(j)}, \bm{y}^{(j)}, \bm{k}^{(j)} \}_{j=1}^N$ be the training dataset, where $\bm{x}^{(j)}$, $\bm{y}^{(j)}$ and $\bm{k}^{(j)}$ denote the clean image, blurry image and blur kernel, respectively, in the $j$th training tuple, and $N$ be the number of training tuples. Our goal is to construct parametric forms with DNNs to approximate the posteriors of both clean image and blur kernel, given only the observed blurry image, under the Bayesian framework. With these parametric variational posteriors trained on $\mathcal{D}$, we can directly infer the posterior distributions of both clean image and blur kernel from a test blurry image. Note that in practical training deblurring datasets \cite{cvpr17_nah_deepdeblur}, for a blurry image, the corresponding ``clean'' image is often obtained by complicated acquisition process, which might not be the exact latent clean image. Besides, the groundtruth blur kernels are also unavailable. Nevertheless, these two issues can be naturally addressed within our Bayesian formulation.
% the simulated clean image, $\bm{k}^{(j)}$ denotes simulated blur kernel,  denotes the corresponding blurred images generated by (\ref{blurmodel}), $N$ is the number of training pair, and the superscript $j$ is the index of training pair. Recall that the blind image deblurring problem can be interpreted as estimating the posterior distribution of latent clean image and blur kernel conditioned on the blurred image. So we can assume that the observed blurred images are generated by some latent random variables and our goal is to infer the latent variables. To this aim, we first formulate a rational Bayesian model of the problem based on the knowledge delivered by the training image pairs.

\subsection{Bayesian Model of Blind Image Deblurring with Data Driven Priors}\label{sec:model}

Let $(\bm{x}, \bm{y}, \bm{k})$, with  $\bm{x} \in \mathbb{R}^{h \times w}$, $\bm{y} \in \mathbb{R}^{h \times w}$, and $\bm{k} \in \mathbb{R}^{h' \times w'} $, be any training tuple in $\mathcal{D}$, where $h \times w$ and $h' \times w'$ are respectively the sizes of image and kernel. As aforementioned, the ``clean'' image $\bm{x}$ and blur kernel $\bm{k}$ might be inaccurate or unavailable, and thus we introduce notations $\bm{z}$ and $\bm{h}$ for the latent clean image and blur kernel, respectively. With these notations, we can build our Bayesian model for blind image deblurring as follows.

\textbf{Likelihood:} Considering the physical blur model \eqref{blurmodel}, we can have the following likelihood for the observed blurry image $\bm{y}$, given the latent clean image $\bm{z}$ and blur kernel $\bm{h}$ \cite{siggraph2006_fergus_removing_shake, tip2016_molina_vi_deblur}:
\begin{equation}\label{blur_likelihood}
	p(\bm{y} | \bm{z}, \bm{h}) = \prod\nolimits_{i=1}^d \mathcal{N}(y_i ; (\bm{h} \ast \bm{z})_i, \sigma^2),
\end{equation}
where $\mathcal{N}(\cdot ; \mu,\sigma^2)$ denotes the Gaussian distribution with mean $\mu$ and variance $\sigma^2$, subscript $i$ denotes the $i$th pixel within the image (by taking the image matrix as a vector) and $d=hw$ for notation convenience. It should be mentioned that, in this work, we treat the variance $\sigma^2$ as a hyper-parameter to reduce the complexity of overall model. Nevertheless, we can also treat it as a to-be-estimated variable and learn its posterior distribution from data as in \cite{nips2019_yue_vdn}. 

To complete the full Bayesian model, we then need introduce priors to $\bm{z}$ and $\bm{h}$. Different from that in traditional optimization based methods, where the priors were specified in a subjective way, we place these priors in a data driven fashion with the aid of training set $\mathcal{D}$, which is detailed in the following. 

\textbf{Data Driven Prior for Clean Image:} As mentioned before, $\bm{x}$ in the training set might not exactly be the latent clean image $\bm{z}$. Nevertheless, it can be expected that $\bm{x}$ is close to $\bm{z}$, and thus provide strong prior information to it. Thus, we can naturally assume the prior distribution of $\bm{z}$ as
\begin{equation}\label{piror_z}
    p(\bm{z}) =  \prod\nolimits_{i=1}^{d} \mathcal{N}(z_i ; x_i, \varepsilon_0^2)   
\end{equation}
where $\varepsilon_0^2$ is a hyper-parameter, which measures the uncertainty of $\bm{z}$. Since typical deblurring datasets are carefully collected, it is expected that $\bm{x}$ is close enough to $\bm{z}$, and we can correspondingly assume a small $\varepsilon_0^2$. In the extreme case that $\varepsilon_0^2$ tends to 0, the prior degenerates to the Dirac distribution with all mass centered at $\bm{x}$. We will empirically analyze the effect of $\varepsilon_0^2$ in our experiments.

\textbf{Data Driven Prior for Blur Kernel:} Specifying a proper prior distribution for the blur kernel is more difficult, since there is a constraint that the sum of all elements in a blur kernel should equal to 1. Fortunately, Dirichlet distribution defined on a simplex well meet this requirement. Its probability density function, with parameter $\bm{\alpha}\!=\!(\alpha _{1}, \dots, \alpha_{M}),\alpha_m\!>\!0,m\!=\!1,\dots,M$ has the following form:
\begin{equation}
	\text{Dir}(\bm{\pi};\bm{\alpha})=\frac{1}{Z(\bm{\alpha})}\prod\nolimits_{m=1}^{M} \pi_{m}^{\alpha _{m}-1},\quad 0<\pi_m<1,\quad m=1,\dots,M,
\end{equation}
where $Z(\bm{\alpha})=\prod_{m=1}^{M}\Gamma(\alpha_m)/\Gamma\big(\sum_{m=1}^{M}\alpha_m\big)$ is the normalizing factor, with $\Gamma(\cdot)$ being the gamma function. Besides, the mean and variance are given by, denoting $\alpha_{0}=\sum_{m=1}^{M}\alpha_m$,
%\begin{equation}\label{dir_mean_variance}
%	{\mathbb{E}}[\pi_{m}] = \frac {\alpha _{m}}{\alpha _{0}}, \quad
%	{\text{Var}}[\pi_{m}] = \frac {\alpha _{m}(\alpha _{0} - \alpha _{m})}{\alpha _{0}^{2}(\alpha _{0} + 1)}.
%\end{equation}
\begin{equation}\label{dir_mean_variance}
{\mathbb{E}}[\pi_{m}] = \alpha _{m}/\alpha _{0}, \quad
{\text{Var}}[\pi_{m}] = \big(\alpha _{m}(\alpha _{0} - \alpha _{m})\big)/\big(\alpha _{0}^{2}(\alpha _{0} + 1)\big).
\end{equation}

With the above knowledge, we can naturally place a Dirichlet prior to the latent blur kernel $\bm{h}$, and the problem is how to set its hyper-parameter. Since $\bm{k}$, if available, in the training set should provide strong information to $\bm{h}$, it is reasonable to assume $\mathbb{E}[\bm{h}]=\bm{k}$. Also, as can be seen from \eqref{dir_mean_variance}, the mean of a Dirichlet distribution remains the same under a scaling transformation of parameters. Therefore, we can introduce the following prior distribution to $\bm{h}$:
\begin{equation}
	p(\bm{h})=\text{Dir}(\bm{h};c\bm{k})\propto\prod\nolimits_{i'=1}^{d'} h_{i'}^{ck_{i'}-1},
	\label{prior_h}
\end{equation}
where $d'=h'w'$ for convenience, and $c>0$ is a scaling factor to be tuned. Although hyper-parameter $c$ does not affect the mean provided $\bm{k}$ being specified, it does determine the variance according to \eqref{dir_mean_variance}, and correspondingly the prior uncertainty of $\bm{h}$. Thus, if $\bm{k}$ is given in the training set, we may set $c$ to a relatively larger value, which results in a smaller prior uncertainty for $\bm{h}$. In contrast, if $\bm{k}$ is unavailable, we can only roughly estimate it from the given $\bm{y}$ and $\bm{x}$, and it is better to set $c$ small to avoid the inaccurate prior information being too strong. %The effect of $c$ is intuitively shown and empirically analyzed in Supplementary Material.

\textbf{Posterior Inference Goal:} Giving both the likelihood and priors, the goal then turns to inferring the posterior distribution of the latent variables $\bm{z}$ and $\bm{h}$ from the observed blurry image $\bm{y}$, which, by the Bayes' rule, can be theoretically formulated as
\begin{equation} \label{true_posterior}
    p(\bm{z}, \bm{h} | \bm{y}) \propto p(\bm{z}, \bm{h}, \bm{y}) = 
    p(\bm{y} | \bm{z}, \bm{h}) p(\bm{z}) p(\bm{h}).    
\end{equation}
Exactly computing this posterior is intractable, especially for a test image out of the training set. Therefore, we turn to constructing the approximate posterior in an amortized way \cite{iclr2014_kingma_vae, icml2014_rezende_2014_vae} with DNNs, and then train the inference networks on the training set. After that, the trained networks can be used to directly infer the posterior of $\bm{z}$ and $\bm{h}$ by input a test blurry image $\bm{y}$.

%\begin{figure}[t]
%    \centering\vspace{-1mm}
%    \includegraphics[center,scale=1.0]{dir_sample_kernel.pdf}
%    \vspace{-4mm}
%    \caption{\small{(a) is the true blur kernel. (b)-(f) are the sample blur kernels with $c$ increasing from left to right}}
%    \label{fig_kernel}
%    \vspace{-4mm}
%\end{figure}

\subsection{Variational Formulations of Posterior}
In order to achieve the goal mentioned in the previous subsection, we first need to construct the variational form of the posterior, denoted as $q(\bm{z}, \bm{h} | \bm{y})$, to approximate the intractable true posterior in \eqref{true_posterior}. Since $p(\bm{z}, \bm{h} | \bm{y})$ is a joint distribution of $\bm{z}$ and $\bm{h}$, there are multiple ways to formulate the variational posterior, and we discuss two possible ones according to the way of inferring $\bm{z}$.

\textbf{Kernel-free Inference Structure for $\bm{z}$:} The basic mean-field assumption for $q(\bm{z}, \bm{h} | \bm{y})$ is
\begin{equation}\label{KF_form}
	q(\bm{z}, \bm{h} | \bm{y}) = q(\bm{z} | \bm{y}) q(\bm{h} | \bm{y}),
\end{equation}
which assumes the conditional independence between $\bm{z}$ and $\bm{h}$ given $\bm{y}$. Then we further specify $q(\bm{z} | \bm{y})$ and $q(\bm{h} | \bm{y})$ with the same families as $q(\bm{z})$ and $q(\bm{h})$, respectively:
\begin{equation}\label{KF_vpost}
	q_{\Psi}(\bm{z} | \bm{y}) = 
	\prod\nolimits_{i=1}^d \mathcal{N}\big(z_i; \mu_i(\bm{y};\Psi), m_i^2(\bm{y};\Psi)\big),\quad q_{\Phi}(\bm{h} | \bm{y}) = \text{Dir}\big(\bm{h}; \bm{\xi}(\bm{y};\Phi)\big),
\end{equation}
where $\Psi$ denotes the parameters of the inference network for $\bm{z}$, which takes $\bm{y}$ as its input and outputs the mean $\mu_i(\bm{y};\Psi)$ and variance $m_i^2(\bm{y};\Psi)$, and thus can be recognized as \emph{Deblur-Net}; similarly, $\Phi$ parameterizes the inference network for $\bm{h}$, which also takes $\bm{y}$ as input while outputs the parameters $\bm{\xi}(\bm{y};\Phi)=\big(\xi_{1}(\bm{y};\Phi),\dots,\xi_{d'}(\bm{y};\Phi)\big)$, and thus can be referred to as \emph{Kernel-Net}.

\textbf{Kernel-dependent Inference Structure for $\bm{z}$:} Another formulation for variational posterior $q(\bm{z}, \bm{h} | \bm{y})$ is obtained by conditional probability rule:
\begin{equation}\label{KD_form}
	q(\bm{z}, \bm{h} | \bm{y}) = q(\bm{z} |\bm{h}, \bm{y}) q(\bm{h} | \bm{y}),
\end{equation}
where the posterior of $\bm{z}$ is also conditioned on $\bm{h}$ in addition to $\bm{y}$. This dependence can be easily achieved by taking both $\bm{h}$ and $\bm{y}$ as the inputs of \emph{Deblur-Net}, leading to following variational form:
\begin{equation}\label{KD_vpost}
	q_{\Psi}(\bm{z} |\bm{h}, \bm{y}) = 
	\prod\nolimits_{i=1}^d \mathcal{N}\big(z_i; \mu_i(\bm{h},\bm{y};\Psi), m_i^2(\bm{h},\bm{y};\Psi)\big).
\end{equation}
This parameterization is rational, since the blur kernel, if accurately estimated, can provide useful information for deblurring an image. The variational form of $q_{\Phi}(\bm{h} | \bm{y})$ remains the same as in \eqref{KF_vpost}.

\emph{Remark:} Mathematically, it is also hold that $q(\bm{z}, \bm{h} | \bm{y}) = q(\bm{h} |\bm{z}, \bm{y})q(\bm{z} | \bm{y})$, which leads to another valid variational posterior formulation. However, considering that our ultimate goal is to deblur $\bm{y}$, this formulation is less attractive, since it requires estimating $\bm{z}$ first before inferring $\bm{h}$.

\subsection{Network Learning with Evidence Lower Bound}
After the variational posterior parameterized by DNNs having be constructed, we need to derive the evidence lower bound (ELBO), which acts as the objective function for learning the network parameters $\Psi$ and $\Phi$. For notation convenience, we simplify $\mu_i(\bm{y};\Psi)$ and $\mu_i(\bm{h},\bm{y};\Psi)$ to $\mu_i$, $m_i^2(\bm{y};\Psi)$ and $m_i^2(\bm{h},\bm{y};\Psi)$ to $m_i^2$, and $\xi_{i'}(\bm{y};\Phi)$ to $\xi_{i'}$ in the following.

Based on conventional decomposition of marginal likelihood in VI, we can get
\begin{equation}
	\log p(\bm{y})=\mathcal{L}(\bm{y};\Psi,\Phi)+D_{KL}\big(q(\bm{z},\bm{h}|\bm{y})\|p(\bm{z},\bm{h}|\bm{y})\big),
\end{equation}
where $D_{KL}(\cdot|\cdot)$ is the KL divergence between two distributions, and $\mathcal{L}(\bm{y};\Psi,\Phi)$ is the ELBO we want to maximize. According to the two variational formulations \eqref{KF_form} and \eqref{KD_form}, we correspondingly have the following two ELBOs (derivations are provided in Appendix \ref{app:elbo_derive}):
\small
\begin{equation}\label{KF_elbo}
	\mathcal{L}_{\text{KF}}(\bm{y};\Psi,\!\Phi)=\mathbb{E}_{q( \bm{z} | \bm{y}) q( \bm{h} | \bm{y})} [ \log p( \bm{y} | \bm{h}, \bm{z} )]- D_{KL}( q(\bm{z} | \bm{y} ) || p( \bm{z}) ) 
	- D_{KL}(q( \bm{h} | \bm{y}) || p(\bm{h})), 
\end{equation}
\normalsize
and
\small
\begin{equation}\label{KD_elbo}
	\mathcal{L}_{\text{KD}}(\bm{y};\Psi,\!\Phi)= \mathbb{E}_{q( \bm{z} | \bm{h}, \bm{y}) q( \bm{h} | \bm{y} )} [ \log p( \bm{y} | \bm{h}, \bm{z})] - \mathbb{E}_{q( \bm{h} | \bm{y})} [ D_{KL}( q( \bm{z} | \bm{h}, \bm{y} ) || p(\bm{z}))]  - D_{KL}(q(\bm{h} | \bm{y}) || p( \bm{h})),
\end{equation}
\normalsize
where we omit the subscripts in $q_{\Psi}(\bm{z} |\bm{h}, \bm{y})$, $q_{\Psi}(\bm{z} |\bm{h}, \bm{y})$ and $q_{\Phi}(\bm{h} |\bm{y})$. The KL divergence terms in the above two equations can then be calculated analytically as follows:
\small
\begin{equation}\label{kl_gaussian}
	D_{KL}(q(\bm{z} | \bm{y}) || p(\bm{z}))=D_{KL}(q(\bm{z} | \bm{h}, \bm{y}) || p(\bm{z})) = 
	\sum_{i=1}^d \left\{\frac{\left(\mu_{i} - x_{i}\right)^{2}}{2 \varepsilon_{0}^{2}}
	+ \frac{1}{2}\left[\frac{m_{i}^{2}}{\varepsilon_{0}^{2}}-\log\frac{m_{i}^{2}}{\varepsilon_{0}^{2}}-1\right]\right\},
\end{equation}
\begin{equation}
\begin{aligned}
D_{KL}(q(\bm{h} | \bm{y}) || p(\bm{h})) = &~\log \Gamma\Big(\sum\nolimits_{i'=1}^{d'}\xi_{i'}\Big) -\log \Gamma\Big(\sum\nolimits_{i'=1}^{d'}ck_{i'}\Big) - \sum\nolimits_{i'=1}^{d'} \log \Gamma(\xi_k) \\
&+ \sum\nolimits_{i'=1}^{d'} \log \Gamma(ck_{i'}) + \sum\nolimits_{i'=1}^{d'} (\xi_{i'} - ck_{i'}) \Big(\psi_0(\xi_{i'}) - \psi_0\Big(\sum\nolimits_{i'=1}^{d'}\xi_{i'}\Big)\Big),
\end{aligned}
\end{equation}
\normalsize
where $\psi_0(\cdot)$ is the digamma function. Note that $D_{KL}(q(\bm{z} | \bm{y}) || p(\bm{z}))$ and $D_{KL}(q(\bm{z} | \bm{y}, \bm{h}) || p(\bm{z}))$ have the same form, since $q(\bm{z} | \bm{y})$ and $q(\bm{z} |\bm{h}, \bm{y})$ both follow Gaussian distributions parameterized by DNNs, while the difference lies in the parametric forms of $\mu_i$ and $m_i^2$ as can be seen in \eqref{KF_vpost} and \eqref{KD_vpost}. The first term in either \eqref{KF_elbo} or \eqref{KD_elbo} can not be analytically calculated due to the intractable integral. Nevertheless, reparameterization trick \cite{iclr2014_kingma_vae, icml2014_rezende_2014_vae, nips2018_figurnov_implicit_trick} can be applied to obtaining an unbiased gradient estimator for it, which has been integrated into modern deep learning frameworks like PyTorch \cite{nips2019_pytorch}.
% and TensorFlow \cite{usenix2016_abadi_tensorflow}.

Given a training dataset $\mathcal{D}$, we now can optimize the network parameters $\Psi$ and $\Phi$ with respect to the ELBO \eqref{KF_elbo} or \eqref{KD_elbo}, over all data in $\mathcal{D}$, which respectively corresponds to the following objectives:
\begin{equation}\label{final_objective}
\max_{\Psi, \Phi} \sum\nolimits_{j=1}^N\mathcal{L}_{\text{KF}}(\bm{y}^{(j)};\Psi,\Phi),\quad \max_{\Psi, \Phi} \sum\nolimits_{j=1}^N\mathcal{L}_{\text{KD}}(\bm{y}^{(j)};\Psi,\Phi).
\end{equation}
\emph{Remark:} The three terms involved in the ELBO, either $\mathcal{L}_{\text{KF}}$ or $\mathcal{L}_{\text{KD}}$, play important roles. Firstly, the second term plays the role of supervising \emph{Deblur-Net} by the clean training image, which acts similarly as the loss functions of most end-to-end trained deblurring DNNs. Secondly, the third term provides supervision information to \emph{Kernel-Net}, which is often ignored in many existing methods. Thirdly, the first term (likelihood), attributed to the physical blur model, includes both \emph{Deblur-Net} and \emph{Kernel-Net}, making them interact with each other, which is also less considered in previous methods. To summarize, with the derived ELBO, \emph{Deblur-Net} and \emph{Kernel-Net} can both be supervised by data-driven priors and interact with each other during the training process.
\begin{figure}[t]
	\centering\vspace{-1mm}
	\includegraphics[width=0.98\textwidth]{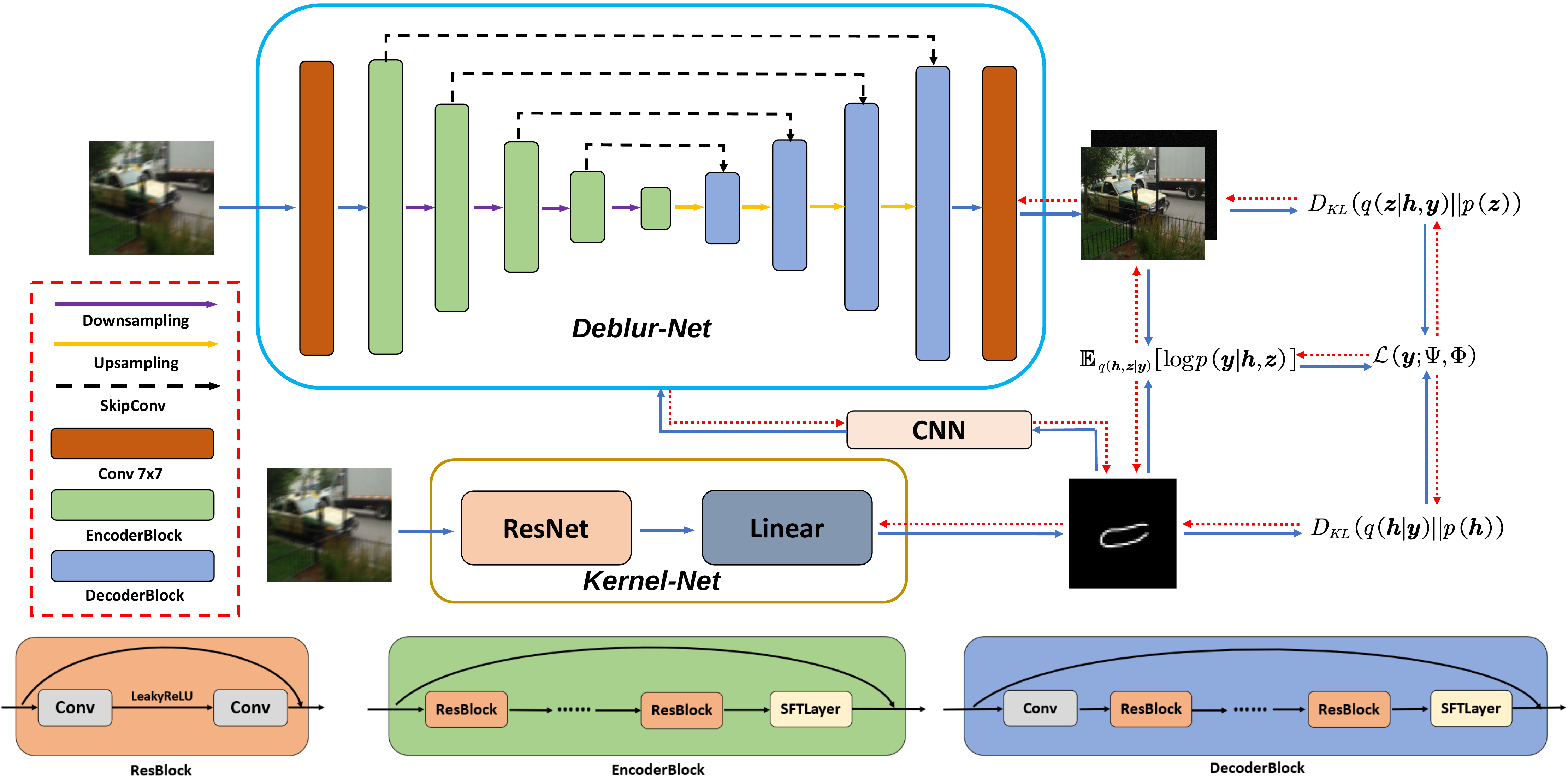}
	\vspace{-2mm}
	\caption{\small{The architecture of the proposed framework with kernel-dependent inference structure. 
	%The blue solid lines denote the forward process, and the red dotted lines mark the gradient flow direction in the BP algorithm.
	}}
	\label{fig_network}
	\vspace{-5mm}
\end{figure}

\subsection{Network Architecture}\label{sec:net_arch}
\textbf{Design of \emph{Deblur-Net}:} For the kernel-free inference structure for $\bm{z}$, we design a relative simple yet effective baseline \emph{Deblur-Net} for evaluation. We adopt the widely used U-Net \cite{miccai2015_unet}, with four scales in the encoder and decoder, to capture multi-scale features of an image, and make some modifications to it. Specifically, we first use the modified ResBlock in \cite{cvpr17_nah_deepdeblur} but replace the ReLU activation with LeakyReLU one and use a small kernel size of 3. Then we stack a series of ResBlocks (the number of ResBlocks varies in our experiments depending on different purposes) to construct the encoding and decoding blocks. The number of channels at all layers is set to 64. The downsample is done by a convolution layer with filter size 3 and stride 2, while the upsample by a transposed convolution layer with filter size 5 and stride 2.

%The kernel-free D-Net takes the blurred image as input and outputs the variational posterior parameters $\bm{\mu}$ and $\bm{m}$. Like many previous studies, we opt for the encoder-decoder architecture (U-Net) \cite{miccai2015_unet} to capture multi-scale information of image. The overall architecture of our proposed D-Net is illustrated in Figure ?. We make some changes in the original U-Net. First, we use the modified ResBlock in \cite{cvpr17_nah_deepdeblur} but repace the ReLU activation with LeakyReLU activation and use a small kernel size of 3. To further improve performance, we stack a series of ResBlock to construct the ConvBlock. The number of channels at all the layers is set to 64. The downsample is done by a convolution layer with filter size 3 and stride 2. The upsample is done by a transposed convolution layer with filter size 5 and stride 2.

For the kernel-dependent inference structure for $\bm{z}$, the \emph{Deblur-Net} should also take the blur kernel, in addition to the blurry image, as its input. To do so, we adopt the SFT layer in \cite{cvpr2019_gu_ikc} to the encoding and decoding blocks. We also use a sub-network to map the blur kernel from matrix to vector. 

\textbf{Design of \emph{Kernel-Net}:} 
%The K-Net takes the blurred image $\bm{y}$ as input and outputs the parameters of variational posterior of blur kernel. Motivated by that Ren \etal \cite{cvpr2020_ren_dip_deblur} simply adopt a fully-connected network to serve as blur kernel generator.
We use a ResNet \cite{cvpr2016_he_resnet} as feature extractor for \emph{Kernel-Net}, followed by a linear transformation layer. Then we apply a softmax layer after that and rescale it to make the final output non-negative and normalized. It should be mentioned that accurately estimating the blur kernel is not a easy task, and there is still a large room to improve the this baseline \emph{Kernel-Net}.

The overall network architecture of the proposed framework with kernel-dependent inference is illustrated in Fig. \ref{fig_network}. For the kernel-free inference, the architecture is almost the same, except that the output of \emph{Kernel-Net} does not be input to \emph{Deblur-Net}.

\subsection{Further Discussions}\label{sec:discussion}
%Before presenting the experimental results, we need to give some further discussions.

\textbf{Non-uniform Blur:} Our deblurring framework is constructed based on the physical blur model \eqref{blurmodel}, which assumes uniform blur kernel across the spatial domain. For non-uniform blurring, we can follow previous studies \cite{cvpr2016_pan_softseg_deblurring, pami2020_pan_physics_gan}, and treat the blur kernel locally uniform within a relatively small image patch.
%However, in real applications, the blur kernel could be non-uniform, which violates our assumption. Fortunately, as mentioned in previous studies \cite{cvpr2016_pan_softseg_deblurring, pami2020_pan_physics_gan}, the non-uniform blur kernel can be regarded locally uniform within a relatively small image patch. 
Since in the training phase, we indeed train the inference networks with patches cropped from images in the training set, the local uniformity of the blur kernel approximately holds, and therefore the proposed framework is also applicable. In the testing phase, the kernel-free inference can be done without any specific modification, while the kernel-dependent inference can be implemented by first performing deblurring on overlapping image patches and then aggregating them to obtain the final result, just like many traditional image restoration methods \cite{ksvd_image, wnnm_j}. However, the above strategy is only a rough approximation, and the application to non-uniform deblurring is still a major limitation of the proposed framework, which should be further investigated in future.

\textbf{Relationship to Kernel-involved Methodologies:} In previous studies, there are mainly two different methodologies in deep learning based deblurring methods that take the kernel information into consideration. The first is represented by Pan \etal \cite{pami2020_pan_physics_gan}, who introduced a loss term motivated by the physical blur model \eqref{blurmodel}. This loss term plays the similar role as the likelihood
%, which is also resulted in by the blur model \eqref{blurmodel}, 
in ELBOs \eqref{KF_elbo} and \eqref{KD_elbo} under our framework. The second is by Kaufman \etal \cite{cvpr2020_kaufman_asnet}, who designed a network to estimate the blur kernel from the blurry image, and input the estimated kernel to the main deblurring network as auxiliary information. This strategy is consistent with the kernel-dependent inference structure for $\bm{z}$ under our framework. 
%as can be seen in Eqs. \eqref{KD_form} and \eqref{KD_vpost}.
Compared with the two existing kernel-involved methodologies, our framework provides a more complete and concise unification, naturally by Bayesian inference principle without any tricky designing, which can not only explain the effectiveness of the existing method from Bayesian perspective, but also facilitate to network training as demonstrated in experiments.

%From these discussions, it can be concluded that the proposed deblurring framework provides a natural unification of two effective methodologies for guiding the deblurring with blur kernel information.
%
%As mentioned before
%As discussed before, our proposed framework can make use of the kernel information by virtue of the Bayesian model. There are two related studies also considering the kernel information with two different methodologies. The first is by Pan \etal \cite{pami2020_pan_physics_gan}, which introduces a loss term motivated by the physical blur model \eqref{blurmodel}. This loss term plays the similar role as the expected log-likelihood term, which is also resulted in by the blur model \eqref{blurmodel}, in ELBOs \eqref{KF_elbo} and \eqref{KD_elbo} under our framework. The second is by Kaufman \etal \cite{cvpr2020_kaufman_asnet}, which designed a network to estimate the blur kernel from the blurry image, and input the estimated kernel to the main deblurring network as side information. This strategy is consistent with the kernel-dependent inference structure for $\bm{z}$ under our framework, as can be seen in Eqs. \eqref{KD_form} and \eqref{KD_vpost}. From these discussions, it can be concluded that the proposed deblurring framework provides a natural unification of two effective methodologies for guiding the deblurring with blur kernel information.

\textbf{Incorporating with Existing Networks:} Though we have designed simple baseline networks in Section \ref{sec:net_arch}, our proposed framework is indeed very general, and can be incorporated with any existing DNNs designed for image deblurring task, as training guidance. As verified in Section \ref{sec:enhance}, the performance can be improved by virtue of the our Bayesian modeling and learning framework.

%\section{Analysis and Discussion}
%It can be seen that the proposed method takes advantages of both model based and deep learning based methods. On one hand, we take advantage of the powerful learning capabilities of deep learning. On the other hand, the proposed method corresponds to the traditional model based methods considering both the image blur model and priors model, which is something that most deep learning methods do not take into account. The work most relevant to our study is Pan \etal \cite{pami2020_pan_physics_gan} and Kaufman \etal \cite{cvpr2020_kaufman_asnet}. Pan \etal \cite{pami2020_pan_physics_gan} introduces the image degradation model in a generative adversarial network for image restoration, which is consistent with our model. The likelihood term in ELBO naturally introduces a image degradation model. However, it does not consider integrating the blur kernel information into the restoration network. The other work \cite{cvpr2020_kaufman_asnet} designs two network for image deblurring and blur kernel predicting respectively, and integrate the blur kernel information to the image deblurring network. However, it does not consider the image degradation model (\ref{blurmodel}), which limits its performance. We should emphasize that our framework is network-independent and it can be adopted in any other different neural network.

\section{Experimental Results}\label{sec:experiments}
In this section, we present the experimental results of the proposed blind image deblurring framework, in comparison with current state-of-the-art methods. We call our method \emph{Variational Bayesian Deblurring Network}, denoted as VBDeblurNet. More results are provided in Appendix \ref{app:more_results}.

\subsection{Experiment Settings}\label{sec:exp_settings}
\textbf{Datasets:} We evaluate the proposed method mainly on three datasets that are widely used for evaluating the deep learning based image deblurring methods, including \emph{Text}
%\footnote{\url{https://www.fit.vutbr.cz/~ihradis/CNN-Deblur/.en}} 
by Hradiš \etal \cite{bmvc15_hradis_cnn_deblur}, \emph{RealBlur}
%\footnote{\url{http://cg.postech.ac.kr/research/RealBlur/}} 
by Rim \etal \cite{eccv2020_rim_realblur} and \emph{GoPro}
%\footnote{\url{https://seungjunnah.github.io/Datasets/gopro.html}} 
by Nah \etal \cite{cvpr17_nah_deepdeblur}. The three datasets represent typical situations we may encounter when applying a deep learning based method. Specifically, the \emph{Text} dataset is composed of blurry-clean-kernel image tuples with uniform blur, 60,000 for training and 1,000 for testing;
%, which is exactly consistent with our model assumptions; 
the \emph{RealBlur}\footnote{We use the \emph{RealBlur-J} subset in experiments.} dataset contains 3,758 blurry-clean image pairs for training and 980 for testing, while the blur kernels are approximately uniform; and the \emph{GoPro} dataset consists of 2,103 blurry-clean training pairs from 22 sequences and 1,111 testing pairs from 11 sequences, with non-uniform blur kernels. As mentioned in Section \ref{sec:model}, on the latter two datasets, the prior information of kernels can be estimated from training patches. 

%The Text dataset consists of 60,000 blurred-clean-kernel image pairs generated by motion blur kernel similar to camera shake and the de-focus blur kernel for training, and provides 100 clean image for test. The image size is 300 $\times$ 300 and the blur kernel size varies from 11 to 27. The RealBlur dataset is a Real-world blur dataset acquired from wide-angle lens and has two sub-datasets: RealBlur-J and RealBlur-R. Each sub-dataset has 3,758 blurred-clean image pairs for training and 980 images for evaluation. The GOPRO dataset contains 22 sequences with 2,103 blurred-clean image pairs for training and 11 sequences with 1,111 blurred-clean image pairs for evaluation. This dataset is usually used for dynamic scene deblurring.
%
%As the dynamic scenes can be approximated by locally uniform blur model \cite{cvpr2016_pan_softseg_deblurring}, we conduct experiment on this dataset. Because the latter two datasets do not provide blur kernels, to make the two datasets adapt to our method, we first estimate a preliminary blur Kernel for next training. The RealBlur dataset is approximately globally uniform blur, so we estimate a blur kernel for an entire image. While the GOPRO dataset is approximately local uniform blur, we randomly crop a small patch for estimating blur kernel. Since all the datasets have been divided into training set and test set, we train on the training set and test on the test set respectively. For the kernel-free mode, we directly input entire image. For the kernel-dependent mode on the GOPRO dataset, we use a small patch input and then  

\textbf{Implementation:} The weights of networks are initialized according to \cite{cvpr2015_he_init}. In each iteration, we randomly crop the entire images to a size of $256 \times 256$ patch for training. The Adam optimizer \cite{iclr2015_kingma_adam} using default parameters is adopted for training, with initial learning rate $1e\text{-}4$, and being reduced to $1e\text{-}5$ as loss becoming stable and finally to $1e\text{-}6$. Without explicitly specified, we set the hyper-parameter $\sigma^2$ in (\ref{blur_likelihood}) to $1e\text{-}5$, $\varepsilon_0^2$ in (\ref{piror_z}) to $1e\text{-}6$, and $c$ in \eqref{prior_h} to $2e4$, respectively. Analyses for the effects of these hyper-parameters are provided in Appendix \ref{app:hyperparam}. 
% We clip gradient by norm to the two networks respectively during the training stage.

\subsection{Model Verification and Ablation Study}
We first construct experiments to verify the effectiveness of the proposed framework with ablation study. Since the \emph{Text} dataset contains the complete and accurate information of blur kernels, we can use it to fully verify the effectiveness of the proposed framework. For a illustration purpose, we use a relatively small \emph{Deblur-Net} with 4 ResBlocks in each encoding or decoding block of the U-Net. We consider four strategies, with different loss functions and inference structures, for obtaining a deblurring network. The first two are training under the supervision of naive mean squared error (MSE), with kernel-free and kernel-dependent inference structures, which are respectively denoted as MSE$_{\text{KF}}$ and MSE$_{\text{KD}}$. The latter two are the proposed framework with kernel-free and kernel-dependent inference structures, denoted as VBDeblurNet$_{\text{KF}}$ and VBDeblurNet$_{\text{KD}}$, respectively. 

The deblurring performance of each strategy, in terms of \emph{peak signal-to-noise-ratio} (PSNR) and \emph{structure similarity} (SSIM) \cite{ssim}, is summarized in Table \ref{ablation_study_table_1}. As can be seen from the table, with the same inference structure, the performance can be largely improved by virtue of the proposed VBDeblurNet framework. Besides, even with the simpler kernel-free inference structure, the proposed VBDeblurNet$_{\text{KF}}$ can outperform MSE$_{\text{KD}}$ that uses kernel information in inference. 
%{\color{red} We also show in Fig. ...}. 
These results evidently verify the effectiveness of the proposed framework. 

\begin{table}[!ht]
	\caption{Deblurring results on \emph{Text} dataset, with different loss functions and inference structures. The best and the second best results are highlighted in bold and Italic bold, respectively.}
	\label{ablation_study_table_1}
	\vspace{-1mm}
	\centering
	\begin{adjustbox}{width=0.6\columnwidth,center}
		\begin{tabular}{c|cccc}
			\toprule
			Method & MSE$_{\text{KF}}$  & VBDeblurNet$_{\text{KF}}$   & MSE$_{\text{KD}}$  & VBDeblurNet$_{\text{KD}}$ \\
			\hline
			PSNR    & 28.70    & \textit{\textbf{29.07}}       & 29.06    & \textbf{29.58} \\
			SSIM    & 0.9845   & \textbf{0.9860}      & 0.9846   & \textit{\textbf{0.9855}} \\
			\bottomrule
		\end{tabular}
	\end{adjustbox}
	\vspace{-4mm}
\end{table} 

%To further demonstrate the effectiveness of our proposed method, we compare the results of simply end-to-end training image delurring networks by MSE loss and using our proposed method. Table \ref{ablation_study_table_1} shows the result. In the table, $a$ and $b$ denotes the result of training the a kernel-free deblurring network and kernel-guided delurring network by MSE loss respectively, the others is result of combining our proposed framework. In can be found that, the proposed method improves the performance. 
%
%To further demonstrate the generality of our proposed method, we compare the results of simply end-to-end training multiple different image delurring networks by MSE loss and using our proposed method. Table \ref{ablation_study_table_2} shows the result. It can be found that the performance of different networks with our framework have been improved.

%%%%%  Text Dasaset
\begin{table}[!t]
	\caption{Comparison of competing methods in terms of PSNR and SSIM on \emph{Text} dataset.}
	\label{text_result}
	\vspace{-2mm}
	\centering
	\begin{adjustbox}{width=\columnwidth,center}
		\begin{tabular}{c|cc|cccc|cc}
			\toprule
			Method   & Xu \etal \cite{cvpr2013_xu_l0norm} & Pan \etal \cite{cvpr2016_pan_dark_channel} & Hradi{\v{s}} \etal \cite{bmvc15_hradis_cnn_deblur}&  Nah \etal \cite{cvpr17_nah_deepdeblur} & Pan \etal \cite{pami2020_pan_physics_gan}  & Kaufman \etal \cite{cvpr2020_kaufman_asnet} & VBDeblurNet$_{\text{KF}}$ & VBDeblurNet$_{\text{KD}}$ \\
			\hline
			PSNR      & 17.52    & 18.47    & 26.53    & 26.81       &28.80      & 27.90  & \emph{\textbf{29.70}}  & \textbf{30.38} \\
			SSIM     & 0.4186   & 0.6127   & 0.9422   & 0.9743      &0.9744     & 0.9604  & \emph{\textbf{0.9862}} & \textbf{0.9872} \\
			\bottomrule
		\end{tabular}
	\end{adjustbox}
	\vspace{-2mm}
\end{table}

\begin{figure}[!t]
	\centering%\vspace{-2mm}
	\includegraphics[width=0.98\textwidth]{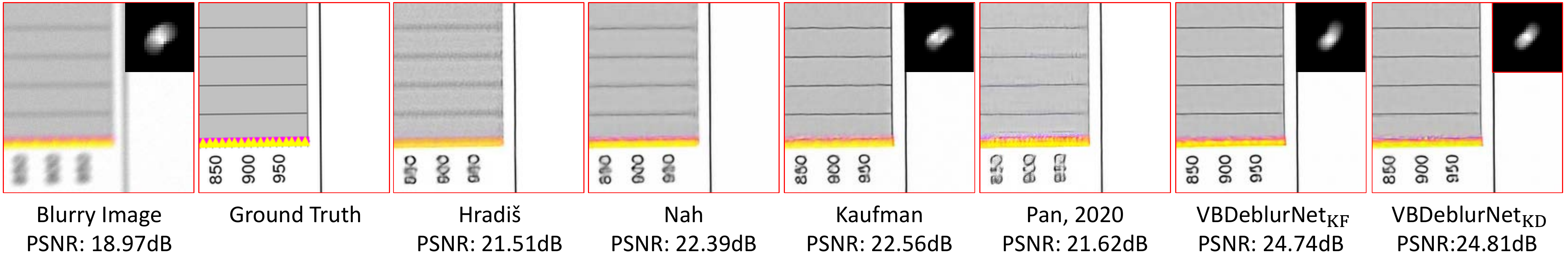}
	\vspace{-2mm}
	\caption{Visual deblurring results of competing methods on \emph{Text} dataset, together with the estimated blur kernels if available.}
	\label{fig_text}
	\vspace{-2mm}
\end{figure}

%%%%% RealBlur dataset
\begin{table}[!t]
	\caption{Comparison of competing methods in terms of PSNR and SSIM on \emph{RealBlur-J} dataset.}
	\label{realblur-j_result}
	\vspace{-1mm}
	\centering
	\begin{adjustbox}{width=\columnwidth,center}
		\begin{tabular}{c|ccc|ccc|cc}
			\toprule
			Method & Hu \etal \cite{cvpr2014_hu_low_light_deblurring} & Xu \etal \cite{cvpr2013_xu_l0norm} & Pan \etal \cite{cvpr2016_pan_dark_channel}& DeblurGAN-v2 \cite{iccv2019_Kupyn_deblurgan_v2} & SRN \cite{cvpr18_tao_srn} & MPRNet \cite{cvpr2021_zamir_mprnet} & VBDeblurNet$_{\text{KF}}$ & VBDeblurNet$_{\text{KD}}$ \\
			\hline
			PSNR    & 26.41   & 27.14    & 27.22    & 29.69     & 31.38     & \emph{\textbf{31.76}}     &\textbf{31.85}  & 31.73\\
			SSIM    & 0.8028  & 0.8303   & 0.7901   & 0.8703    & 0.9091    & \textbf{0.9220}    &\emph{\textbf{0.9132}} & 0.9106\\
			\bottomrule
		\end{tabular}
	\end{adjustbox}
	\vspace{-5mm}
\end{table}

\subsection{Comparison with State-of-the-art Methods}
%{\color{red}{TODO: this subsection needs further revision}}

In this subsection, we compared the proposed method with state-of-the-art methods, including traditional optimization-based methods \cite{cvpr2013_xu_l0norm, cvpr2014_pan_l0text, cvpr2016_pan_dark_channel, cvpr2014_hu_low_light_deblurring} and deep learning-based methods \cite{bmvc15_hradis_cnn_deblur, cvpr17_nah_deepdeblur, pami2020_pan_physics_gan, cvpr2020_ren_dip_deblur, cvpr2018_kupyn_deblurgan, iccv2019_Kupyn_deblurgan_v2, cvpr2021_zamir_mprnet, cvpr18_zhang_dynamic_rnn, cvpr2019_zhang_patch_deblurring}, on the three datasets introduced in Section \ref{sec:exp_settings}. The \emph{Deblur-Net} used here is with 6 Resblocks in each encoding or decoding block, in order to achieve a good performance. We show the quantitative results in terms of two commonly used metrics, PSNR and SSIM, in Tables \ref{text_result}, \ref{realblur-j_result} and \ref{gopro_result}. Note that the competing methods on different datasets are also different. This is because that we want to make sure that each deep learning based method has been trained and tested on the same dataset, by either citing the results from corresponding papers or retraining the models by ourselves. 

It can be seen from Tables \ref{text_result} that, both versions of the proposed framework significantly improve the previous state-of-the-art results on \emph{Text} dataset. This is not surprising, since the structure of the training image tuples in this dataset can be faithfully characterized by our proposed Bayesian generative model, and thus the performance can be naturally benefited by our deep variational inference framework. %{\color{red} Fig. ...} 
On the other two datasets, although the groundtruth blur kernels are unavailable, the benefits of the proposed framework can also be clearly observed from Tables \ref{realblur-j_result} and \ref{gopro_result}. Specifically, the proposed method performs at least comparable with the current state-of-the-art one, i.e., MPRNet, in terms of both PSNR and SSIM, especially that it outperforms MPRNet on \emph{RealBlur-J} dataset in terms of PSNR. The promising performance of MPRNet is not surprising, since authors have made a lot of efforts in network designing. Comparatively, our DNN architecture is much simpler, mainly built on classic structures, such as U-Net and ResBlock, and its performance is mostly attributed to the proposed VBDeblurNet framework. From this perspective, these quantitative results indeed substantiate the effectiveness of the proposed framework. Besides, we can also straightforwardly apply our framework to MPRNet, in order to further pursue better performance. 

%It should be noted that, {\color{red} as summarized in Fig. }, our network is relatively light-weight compared with MPRNet, containing only less than one-third parameters (5.8M v.s. 20.1M). Besides, our method already outperforms the two-stage version of MPRNet on \emph{GoPro} dataset, with about one-half parameters (5.8M v.s. 11.3M). Since our network is with relatively simpler structure and less parameters, these results evidently substantiate the effectiveness of the proposed deep variational deblurring framework. It is also expected that our proposed framework can also be applied to further improving the performance of MPRNet, as suggested in Section \ref{sec:enhance}.

For a more intuitive illustration, we draw the example visual results in Figs. \ref{fig_text}, \ref{fig_realblur} and \ref{fig_gopro} on the three datasets, and more results are provided in Appendix \ref{app:more_results}. Fig. \ref{fig_text} shows visual deblurring results an image in \emph{Text} dataset by different deep learning methods contained in Table \ref{text_result}. As can be seen, the deblurred images by our method have better visual quality, especially for producing sharper edges and digits. Compared the two inference structures of our method, kernel-dependent one unsurprisingly performs better in more accurately recovering digit ``5'', since it makes full use of the kernel information. Besides, the estimated kernels by our method are also close to the ground truth. For the deblurring example on \emph{RealBlur-J} dataset, shown in Fig. \ref{fig_realblur}, our method produces slightly better results than that of MPRNet both in terms of PSNR and visual quality. Visual results on \emph{GoPro} dataset shown in Fig. \ref{fig_gopro} are also very interesting. Specifically, although MPRNet achieves higher PSNR, its visual quality is obviously not the best, especially that it fails to recover the number plate on the car. As a comparison, our VBDeblurNet$_{\text{KD}}$ method can relatively better recover this number plate, especially for digit ``4'' within it.

\begin{figure}[!t]
	\centering
	\includegraphics[width=\textwidth]{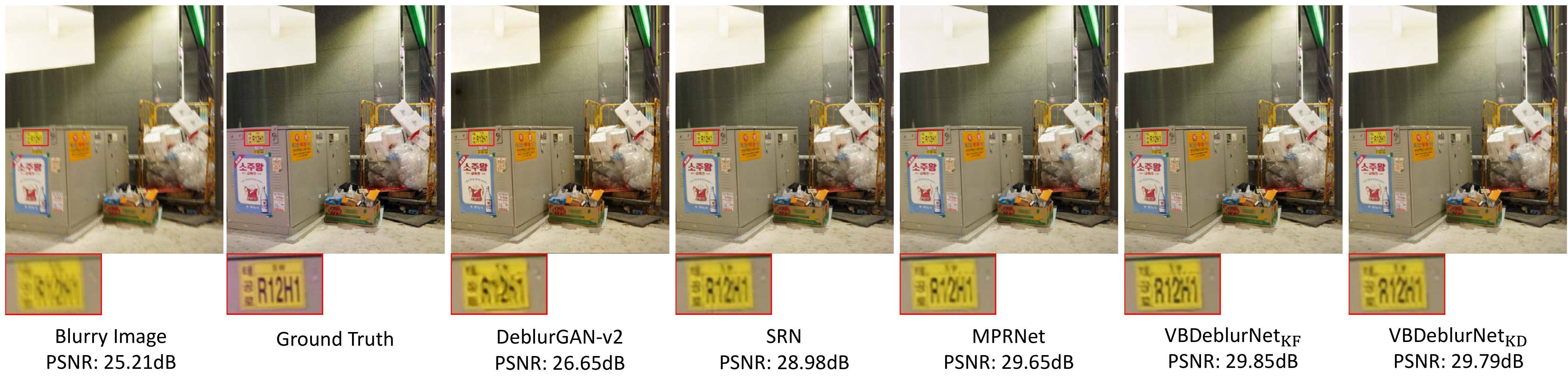}
	\vspace{-6mm}
	\caption{Visual deblurring results of competing methods on \emph{RealBlur-J} dataset.}
	\label{fig_realblur}
	\vspace{-2mm}
\end{figure}

%%%%% GOPRO dataset
\begin{table}[!t]
	\caption{Comparison of competing methods in terms of PSNR and SSIM on \emph{GoPro} dataset.}
	\label{gopro_result}
	\vspace{-2mm}
	\centering
	\begin{adjustbox}{width=\columnwidth,center}
		\begin{tabular}{c|ccccccc|cc}
			\toprule
			Method & Nah \etal \cite{cvpr17_nah_deepdeblur} & DeblurGAN \cite{cvpr2018_kupyn_deblurgan} & SRN \cite{cvpr18_tao_srn}& Gao \etal \cite{cvpr2019_gao_sharing_deblurring} & DBGAN \cite{cvpr2020_zhang_double_gan_deblurring} & Zhang \etal \cite{cvpr2019_zhang_patch_deblurring} & MPRNet \cite{cvpr2021_zamir_mprnet} & VBDeblurNet$_{\text{KF}}$ & VBDeblurNet$_{\text{KD}}$ \\
			\hline
			PSNR    & 29.08  & 28.70 & 30.21    & 30.92     & 31.10             & 31.20       & \textbf{32.66}   & 31.87  & \emph{\textbf{32.03}}   \\
			SSIM    & 0.9135  & 0.858 & 0.9352    & 0.9421    & 0.9424            & 0.940       & \textbf{0.9589}  & 0.9519 & \emph{\textbf{0.9531}}   \\
			\bottomrule
		\end{tabular}
	\end{adjustbox}
	\vspace{-2mm}
\end{table} 

\begin{figure}[!t]
	\centering%\vspace{-1mm}
	\includegraphics[width=0.98\textwidth]{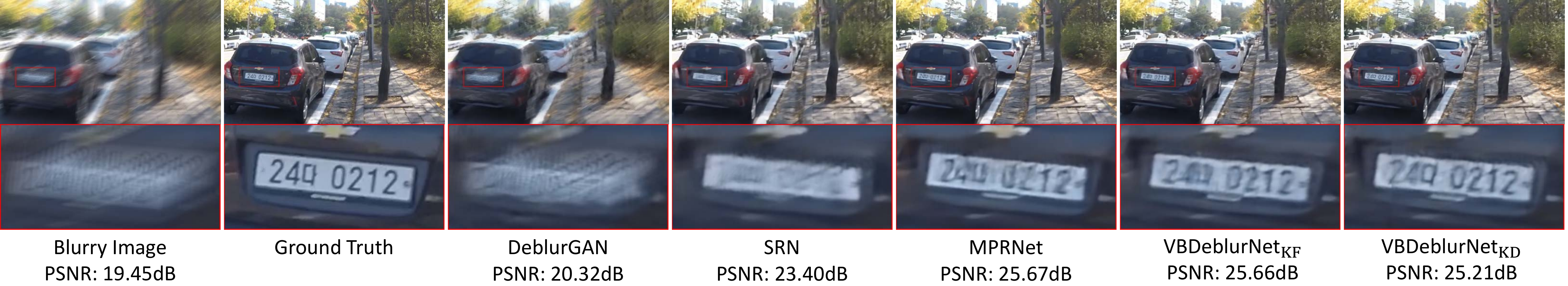}
	\vspace{-2mm}
	\caption{Visual deblurring results of competing methods on \emph{GoPro} dataset.}
	\label{fig_gopro}
	%\vspace{-1mm}
\end{figure}

\subsection{Enhancing Existing Networks by Proposed Framework}\label{sec:enhance}
As discussed in Section \ref{sec:discussion}, the proposed framework can be incorporated with existing DNNs, to improve their deblurring performances. To verify this, we train on the \emph{Text} dataset two existing deblurring networks, i.e., the ones proposed by Nah \etal \cite{cvpr17_nah_deepdeblur}, and Kaufman and Fattal \cite{cvpr2020_kaufman_asnet}, under MSE loss and with our framework, respectively. The former network uses kernel-free inference structure and the latter adopt the kernel-dependent one. The results in terms of PSNR and SSIM are summarized in Table \ref{ablation_study_table_2}. As can be seen from the Table that, the performance of existing networks can be evidently improved by the proposed framework, no matter which types of inference structures they are built on. These improvements can be attributed to two advantages led to by our Bayesian modeling and learning framework. Firstly, the kernel information is naturally included in the training objective as auxiliary supervision. More importantly, \emph{Deblur-Net} and \emph{Kernel-Net} can interact with each other, under the guidance of the likelihood term led to by the physical blur model, to promote the overall training.

\begin{table}[!t]
	\vspace{-2mm}
	\caption{Comparison of existing networks trained with MSE and VBDeblurNet on \emph{Text} dataset.}
	\label{ablation_study_table_2}
	\vspace{-2mm}
	\centering
	\begin{adjustbox}{width=0.75\columnwidth,center}
	\begin{tabular}{c|ccc|ccc}
		\toprule
		\multirow{2}{*}{Method} & \multicolumn{3}{c|}{Nah \etal \cite{cvpr17_nah_deepdeblur}} & \multicolumn{3}{c}{Kaufman \etal \cite{cvpr2020_kaufman_asnet}} \\
		\cline{2-7}
		& MSE  & VBDeblurNet & $\Delta~\uparrow$ &  MSE  & VBDeblurNet & $\Delta~\uparrow$\\
		\hline
		PSNR & 26.81 & 27.03  &  0.22  & 27.90 & 28.10 & 0.20 \\
		SSIM & 0.9743 & 0.9756 & 0.0013 & 0.9607 & 0.9636 & 0.0029 \\
		\bottomrule
	\end{tabular}
	\end{adjustbox}
	\vspace{-4mm}
\end{table}

\section{Conclusion}\label{sec:conclusion} 
We have proposed a new deep variational Bayesian framework, i.e., VBDeblurNet, for blind image deblurring, under which, the approximate posterior of the latent clean image and blur kernel, parameterized by DNNs, can be jointly learned in an amortized inference fashion. By virtue of the fully Bayesian formulation, the inference networks for clean image and blur kernel can both be supervised by data-driven priors, and interact with each other by the likelihood attributed to the physical blur model during training, which thus leads to promising deblurring performance. Comprehensive experiments have verified the effectiveness of the proposed framework, showing that it can not only achieve a promising results with relatively simple networks, but also be able to enhance the deblurring performance of existing DNNs. In the future, we will devote to improving the applicability of VBDeblurNet in non-uniform deblurring.

%\section*{References}
\bibliographystyle{plain}
\bibliography{deblur.bib}

\appendix

\section{Derivation of the Evidence Lower Bounds}\label{app:elbo_derive}
In this section, we derive the evidence lower bounds (ELBOs), or more specifically, Eqs. \eqref{KF_elbo} and \eqref{KD_elbo} in the main text. As presented in the paper, we introduced a variational form of the posterior, denoted as $q(\bm{z},\bm{h}| \bm{y})$, where we omit the subscripts $\Psi, \Phi$, to approximate the intractable true posterior $p(\bm{z},\bm{h}| \bm{y})$. Then we have the following decomposition of the marginal log-likelihood:
\begin{equation}\label{marginal}
\begin{split}
\log p( \bm{y} ) &=\mathbb{E}_{q( \bm{h},\bm{z}|\bm{y} )}\left[ \log p( \bm{y} ) \right]\\
&=\mathbb{E}_{q( \bm{h},\bm{z}|\bm{y} )}\left[ \log \frac{p( \bm{h},\bm{z},\bm{y} )}{p( \bm{h},\bm{z}|\bm{y} )} \right]\\
&=\mathbb{E}_{q( \bm{h},\bm{z}|\bm{y} )}\left[ \log \frac{q( \bm{h},\bm{z}|\bm{y} )}{p( \bm{h},\bm{z}|\bm{y} )}\frac{p( \bm{h},\bm{z},\bm{y} )}{q( \bm{h},\bm{z}|\bm{y} )} \right]\\
&=\mathbb{E}_{q( \bm{h},\bm{z}|\bm{y} )}\left[ \log \frac{p( \bm{h},\bm{z},\bm{y} )}{q( \bm{h},\bm{z}|\bm{y} )} \right] +\mathbb{E}_{q( \bm{h},\bm{z}|\bm{y} )}\left[ \log \frac{q( \bm{h},\bm{z}|\bm{y} )}{p( \bm{h},\bm{z}|\bm{y} )} \right]\\
&=\mathbb{E}_{q( \bm{h},\bm{z}|\bm{y} )}\left[ \log \frac{p( \bm{y}|\bm{h},\bm{z} ) p( \bm{h} ) p( \bm{z} )}{q( \bm{h},\bm{z}|\bm{y} )} \right] +D_{KL}( q( \bm{h},\bm{z}|\bm{y} ) ||p( \bm{h},\bm{z}|\bm{y} ) )\\
&\triangleq\mathcal{L}(\bm{y};\Psi,\!\Phi)+D_{KL}( q( \bm{h},\bm{z}|\bm{y} ) ||p( \bm{h},\bm{z}|\bm{y} ) ).
\end{split}
\end{equation}
The second term in last line of Eq. (\ref{marginal}) is the Kullback-Leibler (KL) divergence between the variational posterior $q(\bm{z},\bm{h}| \bm{y})$ and the true posterior $p(\bm{z},\bm{h}| \bm{y})$, which is non-negative, while the first term is the ELBO, which we want to maximize.

For the kernel-free inference structure of $\bm{z}$, we use the basic mean-field assumption for $q(\bm{z},\bm{h}| \bm{y})$ i.e.,
\begin{equation}
q(\bm{z},\bm{h}| \bm{y}) = q(\bm{z}| \bm{y})\bm{h}| \bm{y}).
\end{equation}
Under this assumption, the ELBO becomes:
\begin{equation}
\begin{split}
\mathcal{L}_{\text{KF}}(\bm{y};\Psi,\!\Phi) &= \mathbb{E}_{q( \bm{h},\bm{z}|\bm{y} )}\left[ \log \frac{p( \bm{y}|\bm{h},\bm{z} ) p( \bm{h} ) p( \bm{z} )}{q( \bm{h},\bm{z}|\bm{y} )} \right] \\
&= \mathbb{E}_{q( \bm{z} | \bm{y}) q( \bm{h} | \bm{y})} [ \log p( \bm{y} | \bm{h}, \bm{z} )] - \mathbb{E}_{q( \bm{z} | \bm{y}) q( \bm{h} | \bm{y})} \left[ \log \frac{q(\bm{h} | \bm{y}) q(\bm{z}|\bm{y} )}{ p( \bm{h} ) p( \bm{z})} \right]\\                                          
&= \mathbb{E}_{q( \bm{z} | \bm{y}) q( \bm{h} | \bm{y})} [ \log p( \bm{y} | \bm{h}, \bm{z} )]- D_{KL}( q(\bm{z} | \bm{y} ) || p( \bm{z}) ) 
- D_{KL}(q( \bm{h} | \bm{y}) || p(\bm{h})).
\end{split}
\end{equation}

For the kernel-dependent inference structure of $\bm{z}$, we use the variational form of $q(\bm{z}, \bm{h} | \bm{y})$ by conditional probability rule:
\begin{equation}
q(\bm{z}, \bm{h} | \bm{y}) = q(\bm{z} |\bm{h}, \bm{y}) q(\bm{h} | \bm{y}).
\end{equation}
Under the setting, the ELBO can be derived as:
\begin{equation}
\begin{split}
\mathcal{L}_{\text{KD}}(\bm{y};\Psi,\!\Phi) 
=&~\mathbb{E}_{q( \bm{h},\bm{z}|\bm{y} )}\left[ \log \frac{p( \bm{y}|\bm{h},\bm{z} ) p( \bm{h} ) p( \bm{z} )}{q( \bm{h},\bm{z}|\bm{y} )} \right] \\
=&~\mathbb{E}_{q( \bm{z} | \bm{y}, \bm{h}) q( \bm{h} | \bm{y})} [ \log p( \bm{y} | \bm{h}, \bm{z} )] - 
\mathbb{E}_{q( \bm{z} | \bm{y}, \bm{h}) q( \bm{h} | \bm{y})} \left[ \log \frac{q(\bm{h} | \bm{y}) q(\bm{z}|\bm{y}, \bm{h} )}{ p( \bm{h} ) p( \bm{z})} \right]\\  
=&~\mathbb{E}_{q( \bm{z} | \bm{y}, \bm{h}) q( \bm{h} | \bm{y})} [ \log p( \bm{y} | \bm{h}, \bm{z} )] - 
\mathbb{E}_{q( \bm{z} | \bm{y}, \bm{h}) q( \bm{h} | \bm{y})} \left[ \log \frac{q(\bm{z}|\bm{y}, \bm{h} )}{p( \bm{z})} \right] \\
&-\mathbb{E}_{q( \bm{z} | \bm{y}, \bm{h}) q( \bm{h} | \bm{y})} \left[ \log \frac{q(\bm{h} | \bm{y})}{ p( \bm{h} )} \right] \nonumber \\
=&~\mathbb{E}_{q( \bm{z} | \bm{h}, \bm{y}) q( \bm{h} | \bm{y} )} [ \log p( \bm{y} | \bm{h}, \bm{z})] - 
\mathbb{E}_{q( \bm{h} | \bm{y})} [ D_{KL}( q( \bm{z} | \bm{h}, \bm{y} ) || p(\bm{z}))]  \\
&- D_{KL}(q(\bm{h} | \bm{y}) || p( \bm{h})).
\end{split}  
\end{equation}

\section{Hyper-parameter Analysis}\label{app:hyperparam}
In the proposed Bayesian models, there are three hyper-parameters to be tuned, i.e., $\sigma^2$ in Eq. (2), $\varepsilon_0^2$ in Eq. (3) and $c$ in Eq. (6) of the main text. Here, we conduct experiments on the \emph{Text} dataset to analyze the effects of these the hyper-parameters. 

\subsection{The Effect of $\sigma^2$}
$\sigma^2$ measures the noise level of the degenerated images, which should be carefully set. As can be seen in Table \ref{sigma_ablation}, the performances are close under when setting $\sigma^2$ to $1e\text{-}4$ and $1e\text{-}5$, and degenerate when setting it too small. This is not surprising, since according to the generation process of the \emph{Text} dataset \cite{bmvc15_hradis_cnn_deblur}, the standard deviation of the true noise comes from $[0,7/255]$, which is consistent with the best setting we have found here. 

\begin{table}[!htbp]
	\caption{Performance under different $\sigma^2$ on the \emph{Text} dataset (kernel-free inference structure with 2 ResBlocks in each encoding and decoding block, and fixing $\varepsilon_0^2$ to $1e\text{-}4$ and $c$ to $1e4$). }
	\label{sigma_ablation}
	\centering
	\begin{tabular}{ccccc}
		\toprule
		$\sigma^2$   & 1e\text{-}4          & 1e\text{-}5 & 1e\text{-}6     &  1e\text{-}7 \\
		\hline
		PSNR         & 27.51       & 27.52       & 27.37           & 27.03        \\
		SSIM         & 0.9796      & 0.9796      & 0.9781          & 0.9657       \\
		\bottomrule
	\end{tabular}
\end{table}

\subsection{The Effect of $\varepsilon_0^2$}
$\varepsilon_0^2$ determines how much does the desired latent clean image $\bm{z}$ depend
on the ``clean'' training image $\bm{x}$. It can be observed from Table \ref{epsilon_ablation} that, as $\varepsilon_0^2$ decreasing, the performance first increases and then decreases. This can be naturally explained. Since in the \emph{Text} dataset, the clean training image is indeed the ground truth one, it is better to use a small $\varepsilon_0^2$. However, if setting $\varepsilon_0^2$ too small, the risk of overfitting increases. 

\begin{table}[!htbp]
	\caption{Performance under different $\varepsilon_0^2$ on the Text dataset (kernel-free inference structure with 2 ResBlocks in each encoding and decoding block, and fixing $\sigma^2$ to $1e\text{-}6$ and $c$ to $2e4$).}
	\label{epsilon_ablation}
	\centering
	\begin{tabular}{ccccc}
		\toprule
		$\varepsilon_0^2$ & 1e\text{-}4 & 1e\text{-}5 & 1e\text{-}6     & 1e\text{-}7       \\
		\hline
		PSNR              & 26.75       & 27.00       & 27.26   & 27.09             \\
		SSIM              & 0.9638      & 0.9761      & 0.9783  & 0.9779         \\
		\bottomrule
	\end{tabular}
\end{table}

\subsubsection{The Effect of $c$}
\begin{figure}[t]
	\centering
	\includegraphics[center,scale=1.0]{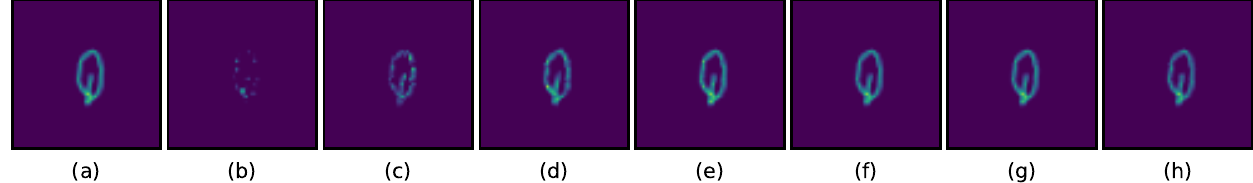}
	\caption{Illustration for the effect of hyper-parameter $c$. (a): the true blur kernel; (b)-(f): blur kernels sampled with $c$ being set to $1e1$, $1e2$, $1e3$, $5e3$, $1e4$, $2e4$ and $3e4$, respectively.}
	\label{fig_kernel}
\end{figure}

$c$ reflects our confidence about the prior information of kernel. Its effect to the kernels sampled from the Dirichlet distribution can be intuitively observed in Fig. \ref{fig_kernel}. Specifically, small $c$ leads to large uncertainty of the sampled kernel, and vice versa. Besides, if $c$ is too small, the sampled kernel will largely deviate from the ground truth one, and therefore relatively larger $c$ is preferred. Also, since in \emph{Text} dataset, the kernels are manually synthesize, we can strongly rely on them, and correspondingly set $c$ to a large value. This is verified by the results in Table \ref{c_ablation}, that the performance increases as $c$ tends to large. Note that we do not further increase $c$ due to the numerical instability.

\begin{table}[!htbp]
	\caption{Performance under different $c$ on the Text dataset (kernel-dependent inference structure with 4 ResBlocks in each encoding and decoding block, and fixing $\sigma^2$ to $1e\text{-}5$ and $\varepsilon_0^2$ to $1e\text{-}6$).}
	\label{c_ablation}
	%\vspace{-2mm}
	\centering
	\begin{tabular}{ccccc}
		\toprule
		$c$  & 5e3          & 1e4      &  2e4           & 3e4              \\
		\midrule
		PSNR        & 29.32        & 29.45       & 29.48           & 29.58         \\
		SSIM        & 0.9857            & 0.9859      & 0.9859          & 0.9856        \\
		\bottomrule
	\end{tabular}
\end{table}

\section{More Visual Results}\label{app:more_results}
\begin{figure}[htbp]
	\centering
	\vspace{-2mm}
	\includegraphics[width=\textwidth]{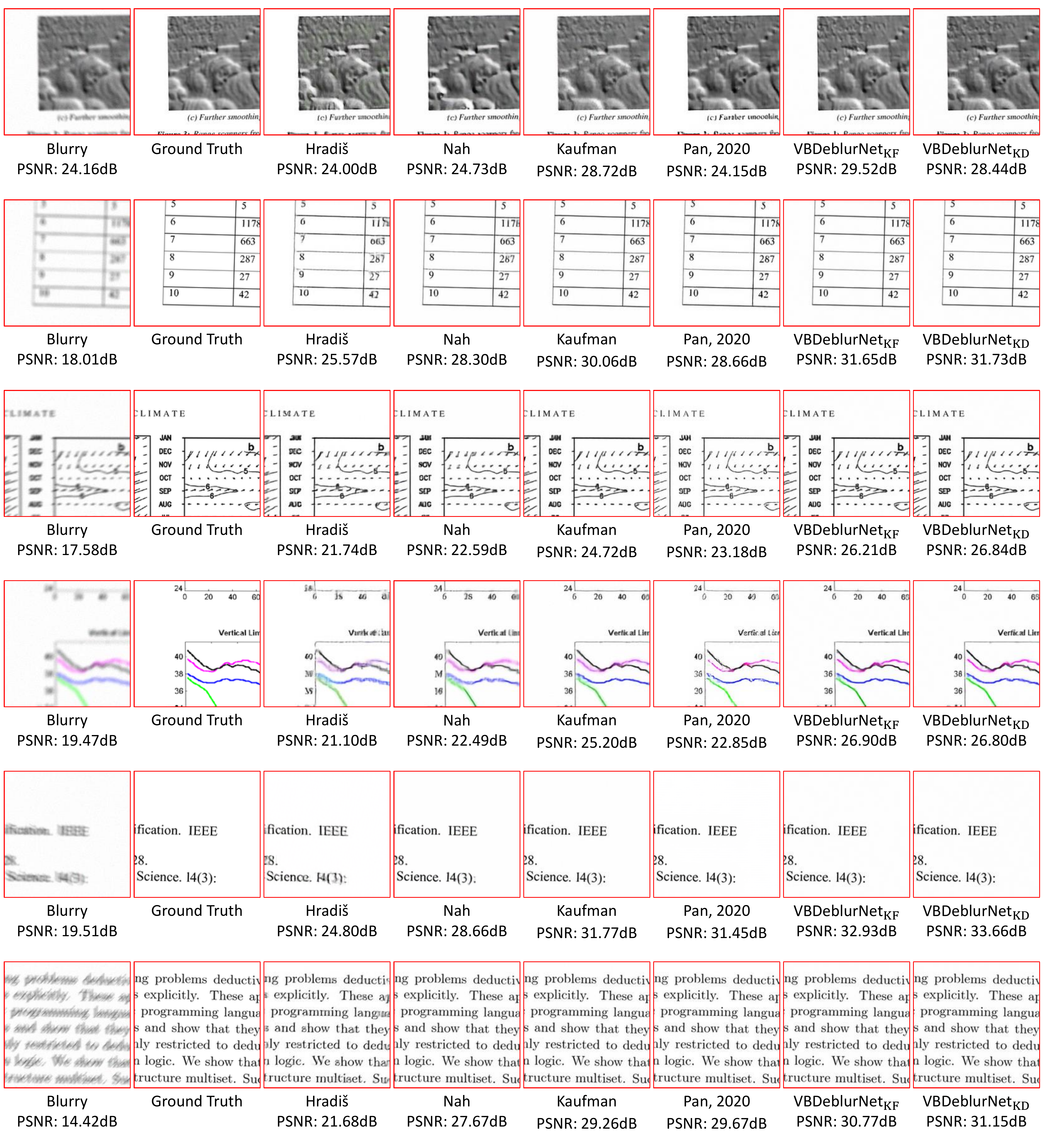}
	\vspace{-6mm}
	\caption{More Visual deblurring results of competing methods on \emph{Text} dataset.}
	\label{fig_text_more}
	\vspace{-2mm}
\end{figure}

\begin{figure}[htbp]
	\centering
	\vspace{-2mm}
	\includegraphics[width=\textwidth]{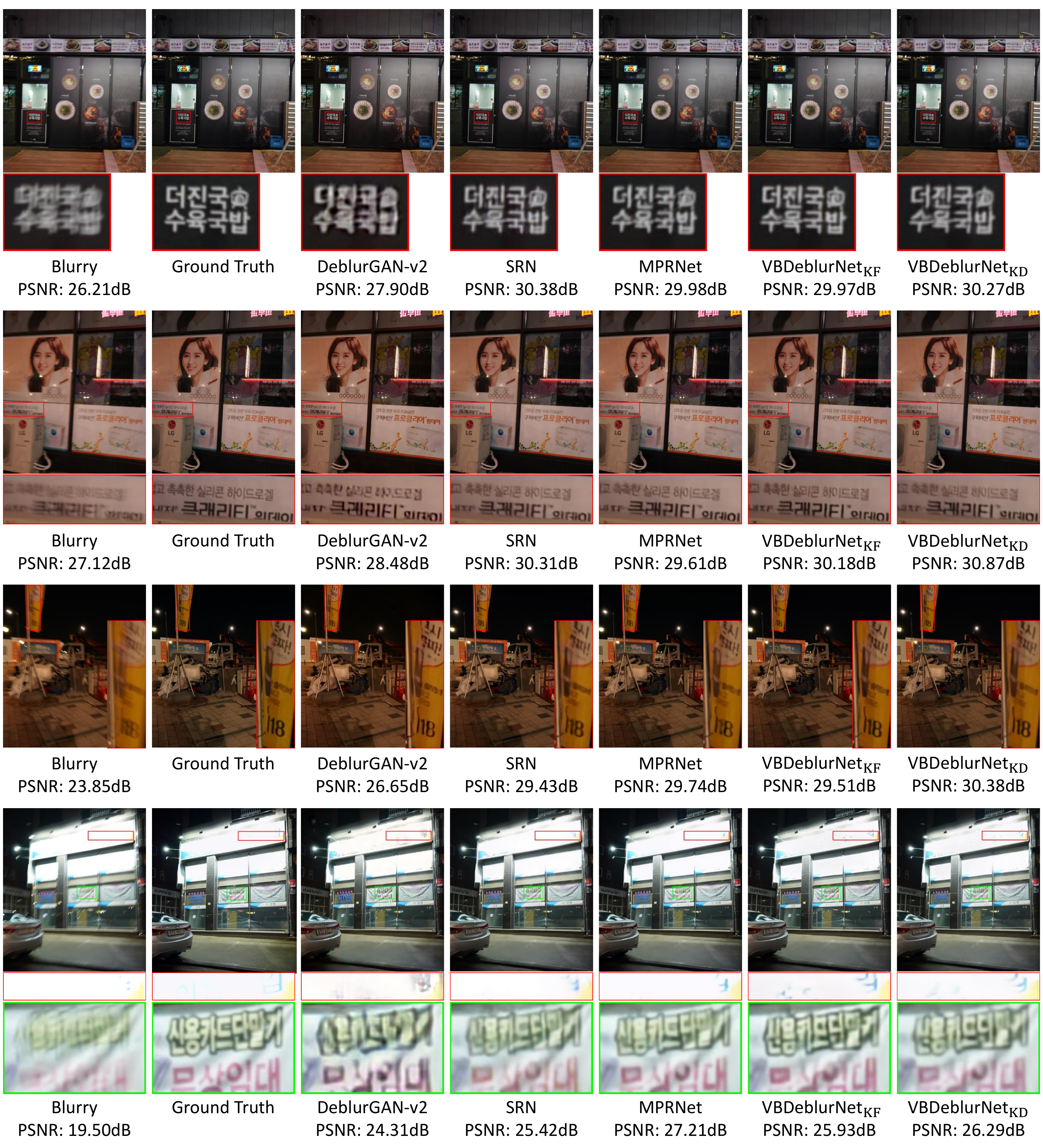}
	\vspace{-6mm}
	\caption{More Visual deblurring results of competing methods on \emph{RealBlur-J} dataset.}
	\label{fig_gopro_more}
	\vspace{-2mm}
\end{figure}

\begin{figure}[htbp]
	\centering
	\vspace{-2mm}
	\includegraphics[width=\textwidth]{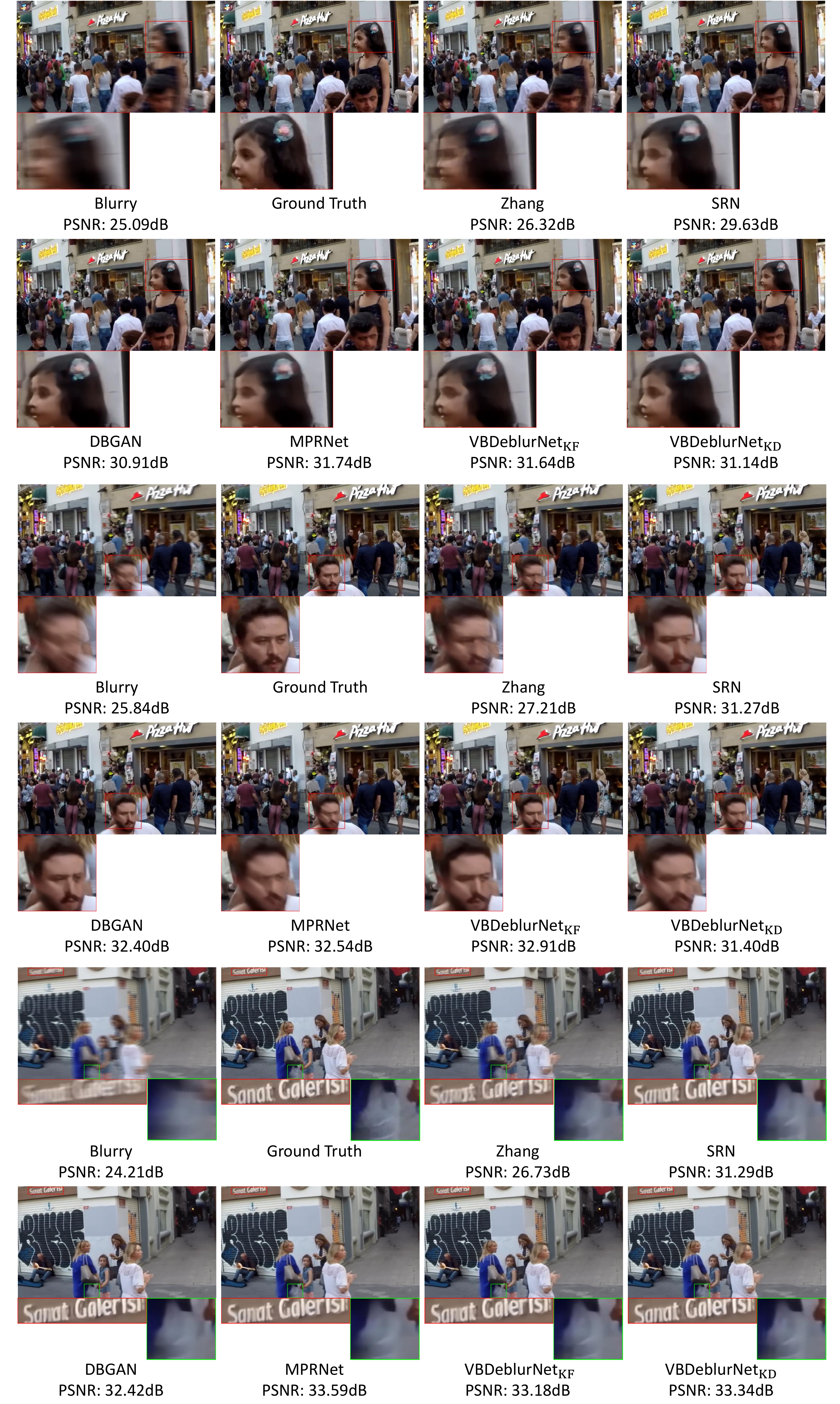}
	\vspace{-6mm}
	\caption{More Visual deblurring results of competing methods on \emph{GoPro} dataset.}
	\label{fig_realblur_more}
	\vspace{-2mm}
\end{figure}

In this Section, we provide more visual results on the three datasets we considered in experiments, as shown in Figs. \ref{fig_text_more}, \ref{fig_realblur_more} and \ref{fig_gopro_more}, respectively. These results further substantiate the effectiveness of the proposed method.

\end{document}